\documentclass{PoS}
\usepackage{amsmath}
\usepackage{amssymb}
\usepackage{multirow}
\usepackage{graphicx}
\usepackage{float}

\title{
  Locality of the overlap-Dirac operator on topology-fixed gauge
  configurations 
}

\ShortTitle{Locality of the overlap-Dirac operator}

\author{\speaker{Yong-Gwi Cho} \\
 Graduate School of Pure and Applied Sciences, University of Tsukuba\\
 Tsukuba, Ibaraki 305-8571, Japan\\
  E-mail: \email{cho@ccs.tsukuba.ac.jp}}

\author{{Shoji Hashimoto}\\
  High Energy Accelerator Research Organization (KEK) and\\
  School of High Energy Accelerator Science, The Graduate University
  for Advanced Studies\\
  Tsukuba 305-0801.\\
  E-mail: \email{shoji.hashimoto@kek.jp}
}

\abstract{
  We investigate the locality property of the overlap-Dirac operator
  on gauge configurations generated with extra Wilson fermions. By
  such extra terms we expect that the structure of the Aoki phase
  would change drastically. In particular, we study the possibility of
  defining the overlap-Dirac operator in the strong coupling regime
  keeping its exponential locality.
}

\FullConference{The 30th International Symposium on Lattice Field Theory\\
		 June 24--29,  2012\\
		 Cairns, Australia}

\begin{document}

\section{Introduction}

Locality of the overlap-Dirac operator is not obvious, since its
definition \cite{Neuberger:1998wv} 
\begin{equation}
  aD_{ov} = 1+\gamma_{5}\frac{H_{W}}{\left|H_{W}\right|}\label{eq:1},
\end{equation}
includes an operator $H_W$ in the denominator.
Here, $H_{W}$ is the hermitian Wilson-Dirac operator
$H_{W}=\gamma_{5}D_{W}$, that is used as a kernel to construct $D_{ov}$.
Superficially, if the eigenvalue spectrum of $H_{w}$ contains
near-zero modes, the overlap-Dirac operator may violate locality.

The locality is known to be satisfied at weak couplings.
To be specific, on background gauge configurations satisfying some
{\it smoothness} condition, one can show that $D_{ov}$ is
exponentially localized \cite{Hernandez:1998et}, {\it i.e.}
$|(D_{ov})_{xy}|\le\exp(-|x-y|/\ell)$
with a localization length $\ell$.
This condition is however too strong for practical setup used in
present lattice QCD simulations;
numerical tests are necessary for more realistic cases.

Golterman and Shamir conjectured that the overlap-Dirac operator
defined outside of the Aoki phase is local 
\cite{Golterman:2003qe}.
Contrary to the original argument \cite{Aoki:1983qi}, the Aoki phase in this case
defined by the profile of near-zero modes:
inside the Aoki phase the near-zero modes are extended in space, while
they are localized outside.
The value of eigenvalue $\lambda_c$ above which the eigenmodes are
extended is called the mobility edge borrowing the terminology of
condensed matter physics.
Then, the localization length of the overlap-Dirac operator is
determined either by $1/\lambda_c$ or $\ell(\lambda)$
($|\lambda|<\lambda_c$), where $\ell(\lambda)$ is the localization
length of the individual low-lying modes.
Thus, the question of the locality crucially depends on the background
gauge field.

The origin of the near-zero modes of $|H_W|$ is the
{\it roughness} of the gauge configuration.
A simple analytic example is given in \cite{Berruto:2000fx}.
Therefore, the localization length is expected to increase toward
strong couplings or coarse lattices,
and the definition of the overlap-Dirac operator becomes more
difficult. 
To avoid this problem one may introduce additional terms to the
lattice action, such as those proposed in \cite{Fukaya:2006vs}.
They consist of two flavors of heavy Wilson fermions and their
associated ghosts carrying a twisted mass term
\begin{equation}
  S_{ex} = 
  \sum_x \bar{\chi}\left(x\right) 
  D_{W}\left(m_{0}\right) \chi\left(x\right)
  +
  \sum_x \bar{\phi}\left(x\right)
  \left[D_{W}\left(m_{0}\right)+i\mu\gamma_{5}\tau_{3}\right]\phi\left(x\right)
  \label{eq:3},
\end{equation}
where $\chi$ denotes Wilson fermions with a negative mass $m_{0}$. 
The second term represents bosonic fields to cancel the bulk of the
effects of Wilson fermions.
In fact, the action generates a suppression factor 
\begin{equation}
  \det\left[\frac{H_{W}\left(m_{0}\right)^{2}}{H_{W}\left(m_{0}\right)^{2}+\mu^{2}}\right]\label{eq:4}
\end{equation}
in the partition function. 
Then, the gauge configuration with small eigenvalues of $H_W(m_0)$
lower than $\mu$ is suppressed,
and the potentially dangerous near-zero modes disappear from the
eigenvalue spectrum. 
Since the low-lying eigenvalue can never cross zero due to the
suppression factor $\det H_W(m_0)^2$, global topology of the gauge
field configuration does not change under continuous deformations.
The simulations are thus confined in a given topological sector with
this action.

In this work, we investigate the spectrum of $H_W$ and the locality
of $D_{ov}$ in the strong coupling regime with or without the extra
Wilson fermion terms.
The purpose of the study is to explore the possibility to use the
overlap fermion at coarser lattices than currently available ones.

\section{Eigenvalue distribution}

\begin{table}[tb]
\begin{center}
\begin{tabular}{|c|ccccc|cccc|}
\hline 
$\mu$ & \multicolumn{5}{c|}{0} & \multicolumn{4}{c|}{0.2}\\
\hline 
$\beta$ & 5.95 & 5.83 & 5.63 & 5.50 &5.43 & 5.78 & 5.68 & 5.48 & 5.28\\
\hline 
$a$(fm) & 0.10 & 0.12 & 0.15 & 0.27 & - & 0.10 & 0.12 & 0.14 & 0.27\\
\hline 
\end{tabular}
\end{center}
\caption{Lattices parameters}
\label{Table1}
\end{table}

In this study, we use quenched lattices (no dynamical {\it light}
fermions) at the $\beta$ values listed in Table~\ref{Table1}.
Lattice size is $16^3\times 32$ and the gauge action is the standard
Wilson gauge action.
We consider the lattices with and without the low-mode suppression
term (\ref{eq:4}).
In the Table, the lattices without that term are denoted by $\mu=0$ as
the extra factor cancels in this case.
The large negative mass $am_0$ is $-1.6$, and the number of gauge
configurations studied is 10 for each parameter.
Lattice spacing $a$ is estimated using the Sommer scale $r_0$
extracted from the static quark potential.

\begin{figure}[tb]
  \begin{center}
   \includegraphics[width=10cm]{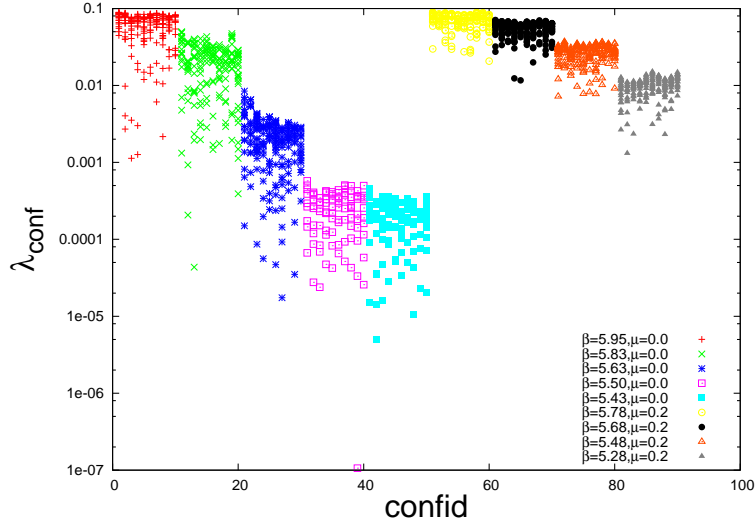}
  \end{center}
  \caption{
    Near-zero eigenvalue distribution of $|H_W|$ plotted in the
    logarithmic scale.
    The five clusters on the left show the results for the standard gauge
    action while the right four clusters are those with the low-mode
    suppression term.
  }
  \label{fig1}
\end{figure}

We calculate 20--40 lowest eigenvalues of $|H_W|$ on these lattices.
Figure~\ref{fig1} shows those near-zero eigenvalue distribution in a
logarithmic scale.
It is clear that the number of low-modes increases on coarser (or
smaller $\beta$) lattices.
With the low-mode suppression term (four right clusters), they are
indeed highly suppressed.
Compared at similar lattice spacings, the lowest eigenvalue is 1--2
orders of magnitude higher.

According to the Banks-Casher relation \cite{Banks:1979yr}, the
absence of the low-lying modes implies that the flavor-parity symmetry
broken phase as defined by Aoki is not entered for these lattices.
This is the effect of the low-mode suppressing term, though the gap is
rather small at coarser lattices.
The practical question is then how large the localization length is
for these coarse lattices, which is addressed in the next section.

\section{Localization of eigenmodes}
In order to investigate the locality of $D_{ov}$ we look at the
spatial profile of low-lying modes.
For each eigenmodes $\phi_i(x)$ of $H_{W}$, we define
$\rho_i(x)$ and $f_i(r)$ as 
\begin{eqnarray}
  \rho_{i} & =  &
  \phi_{i}^{\dagger}(x)\phi_{i}(x),
  \;\;
  \rho_{i}(x_{0}) =
  \underset{x}{max}\left\{ \rho_{i}\left(x\right)\right\} ,
  \\
  f_{i}\left(r\right) & = & 
  \left\{ \rho_{i}\left(r\right)|r=\left|x-x_{0}\right|\right\}
\end{eqnarray}
following to \cite{Yamada:2006fr}.
Namely $\rho_i(x)$ is the strength of the mode, and 
$f_i(r)$ represents the profile of that mode as a function of
the distance from the position where $\rho_i(x)$ has a maximum.
In calculating $f_i(r)$, different orientations giving the same $r$
are averaged.

\begin{figure}[tb]
  \begin{center}
      \includegraphics[scale=0.45]{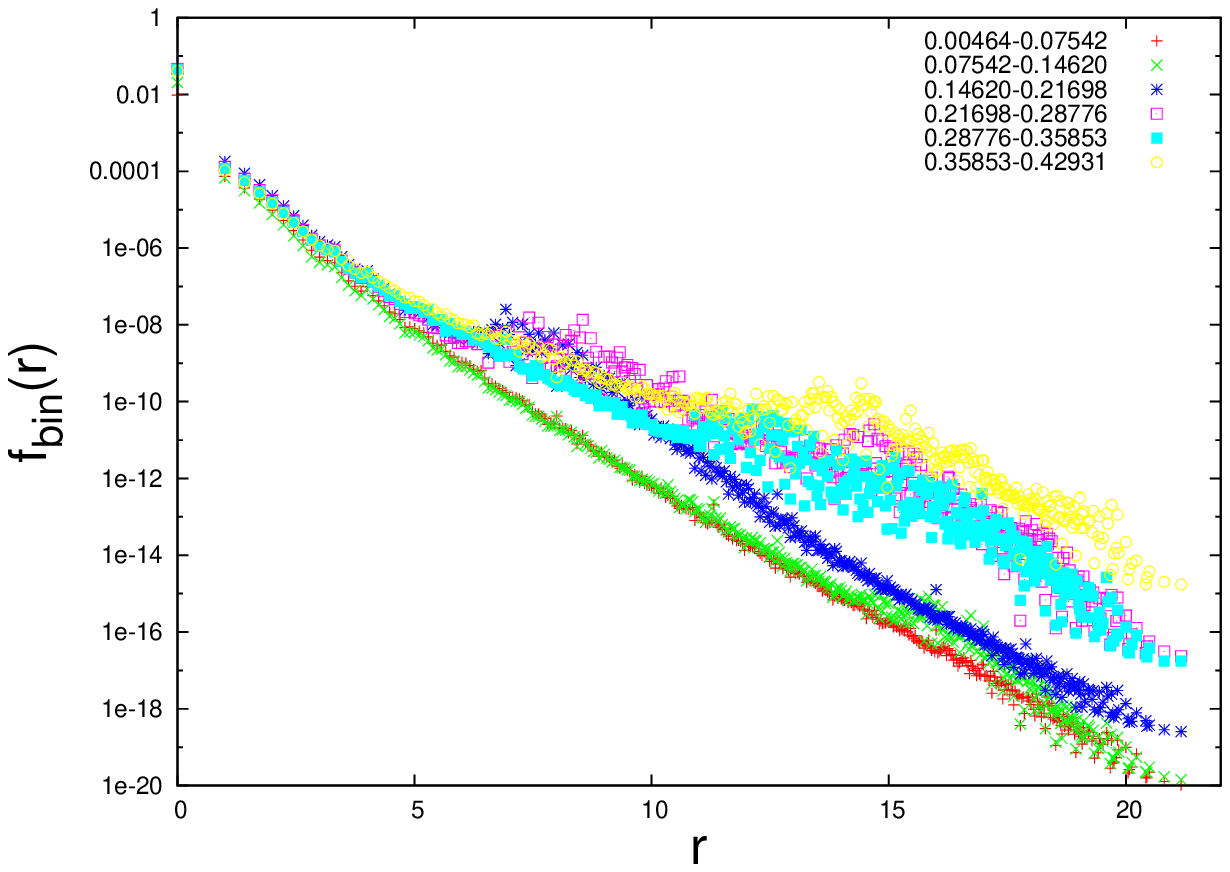}
      \includegraphics[scale=0.45]{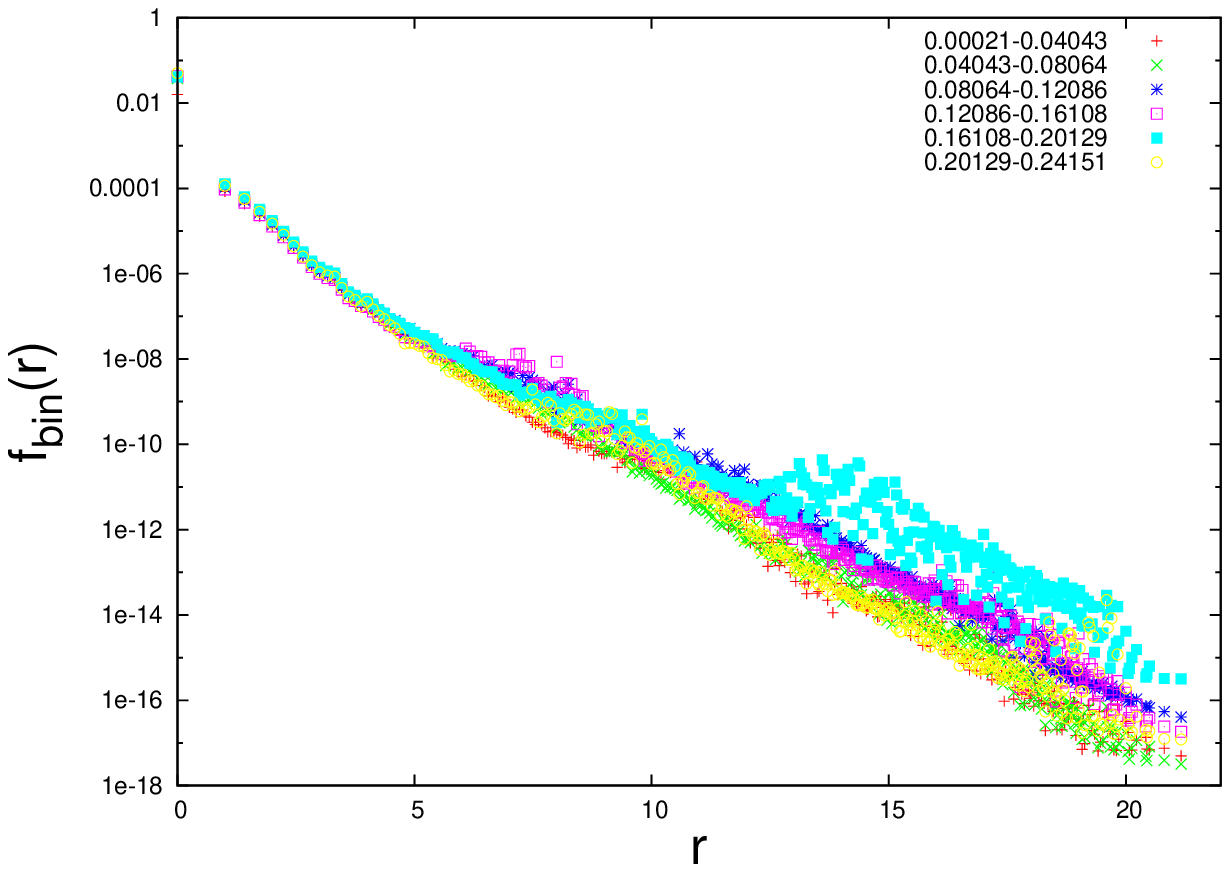}\\
      \includegraphics[scale=0.45]{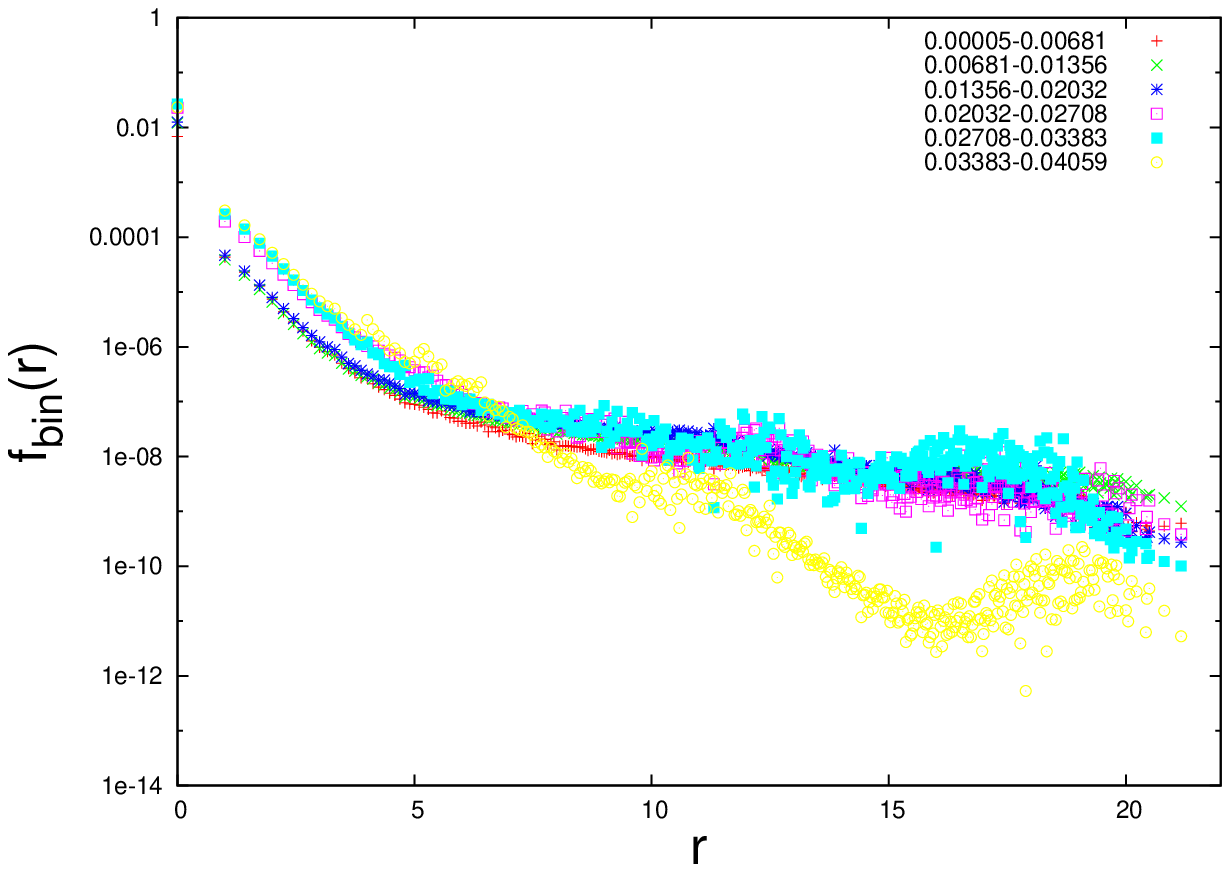}
      \includegraphics[scale=0.45]{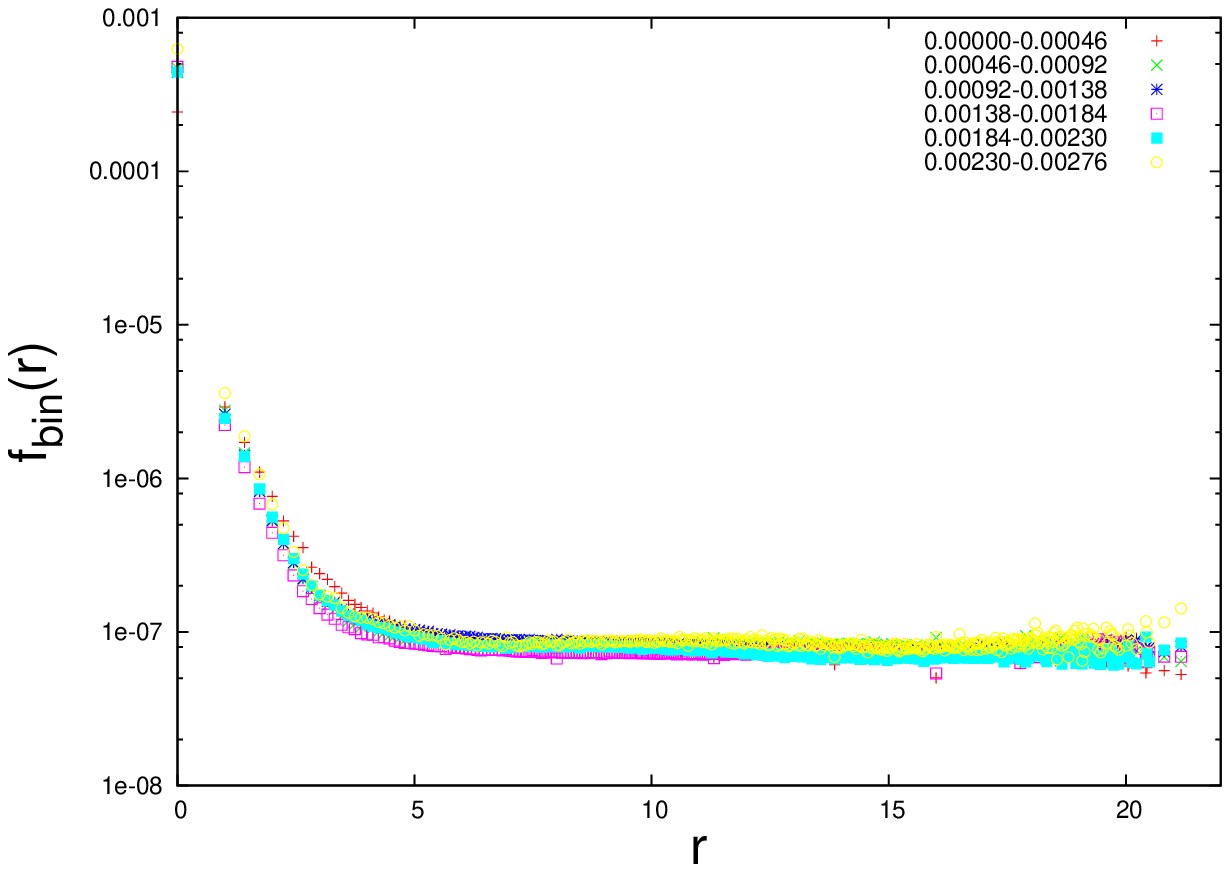}\\
      \includegraphics[scale=0.45]{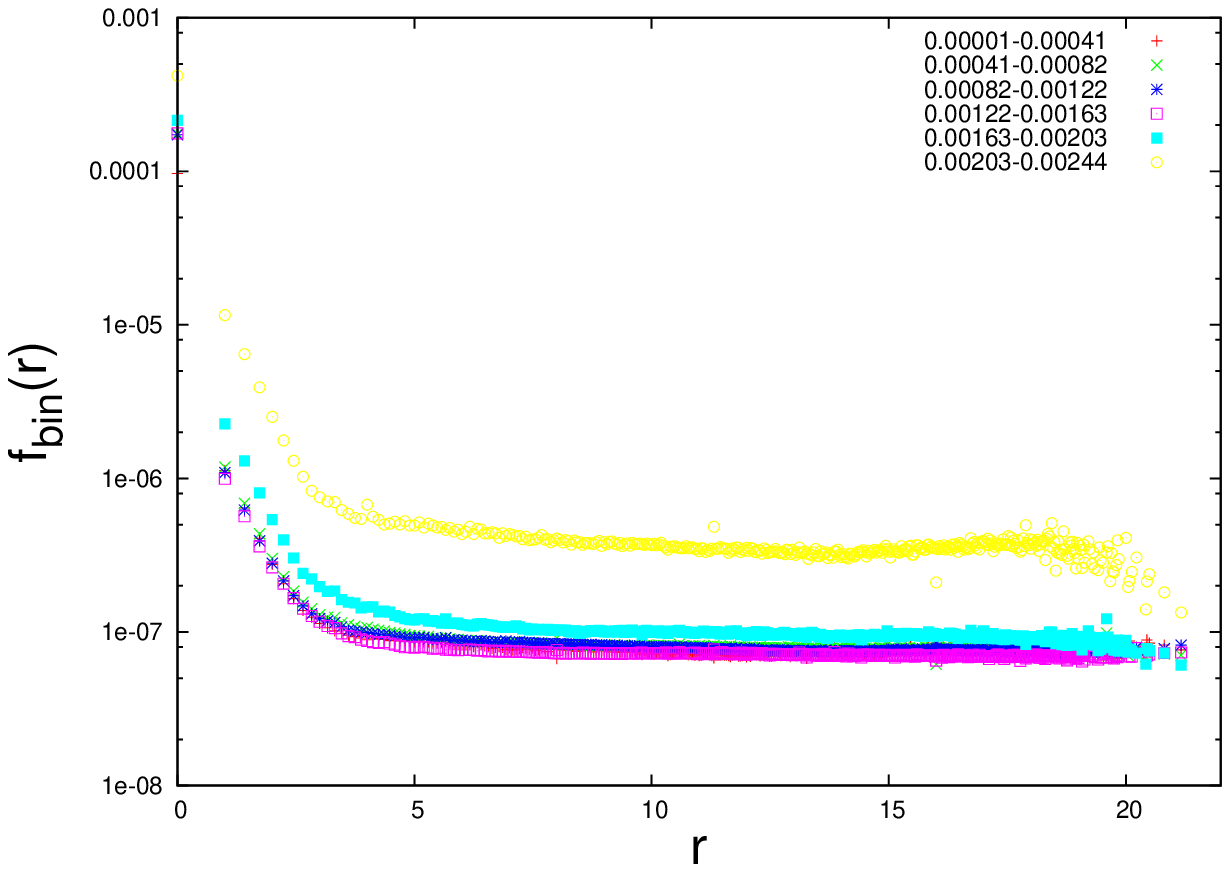}
  \end{center}
  \caption{
    $f_{bin}(r)$ on the quenched gauge configurations without the
    low-mode suppressing term ($\mu=0.0$).
    The data are shown for 
    $\beta=5.95$ (top left), $\beta=5.83$ (top right),$\beta=5.63$
    (middle left), $\beta=5.50$ (middle right), and
    $\beta=5.43$ (bottom).
    The eigenmodes are binned in different ranges of the eigenvalues as
    indicated in the legend of each plot.
  } 
  \label{fig2}
\end{figure}

In Figure \ref{fig2}, we plot $f_i(r)$ averaged over configurations
after binning the eigenmodes in different ranges of their eigenvalues,
that we call $f_{bin}(r)$.
At the $\beta$ values above 5.63, we find a clear fall-off of the
eigenmodes as a function of $r$.
That is true even at the highest bin we measured.
This indicates that $D_{ov}$ constructed on these gauge configurations
is local with the length controlled by the fall-off of these low-lying
modes.

At $\beta$=5.50 and 5.43, on the other hand, we find that $f_{bin}(r)$
becomes flat beyond $r\simeq 5$.
It means that the system is already in the Aoki phase.
This results is consistent with results of \cite{Golterman:2005fe},
where the mobility edge falls down to zero at $\beta=5.5$ and
$am_0=-1.5$.

\begin{figure}[tb]
  \begin{center}
    \includegraphics[scale=0.45]{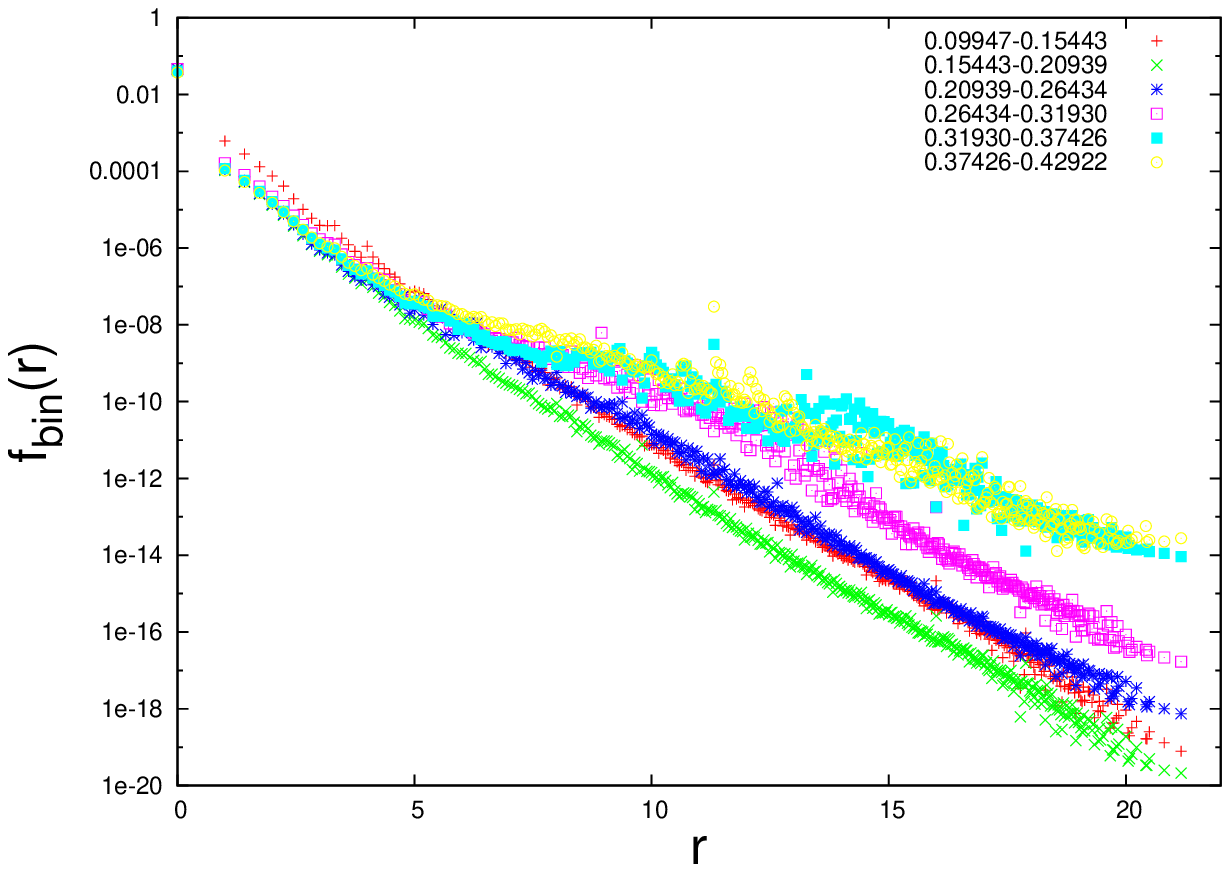}
    \includegraphics[scale=0.45]{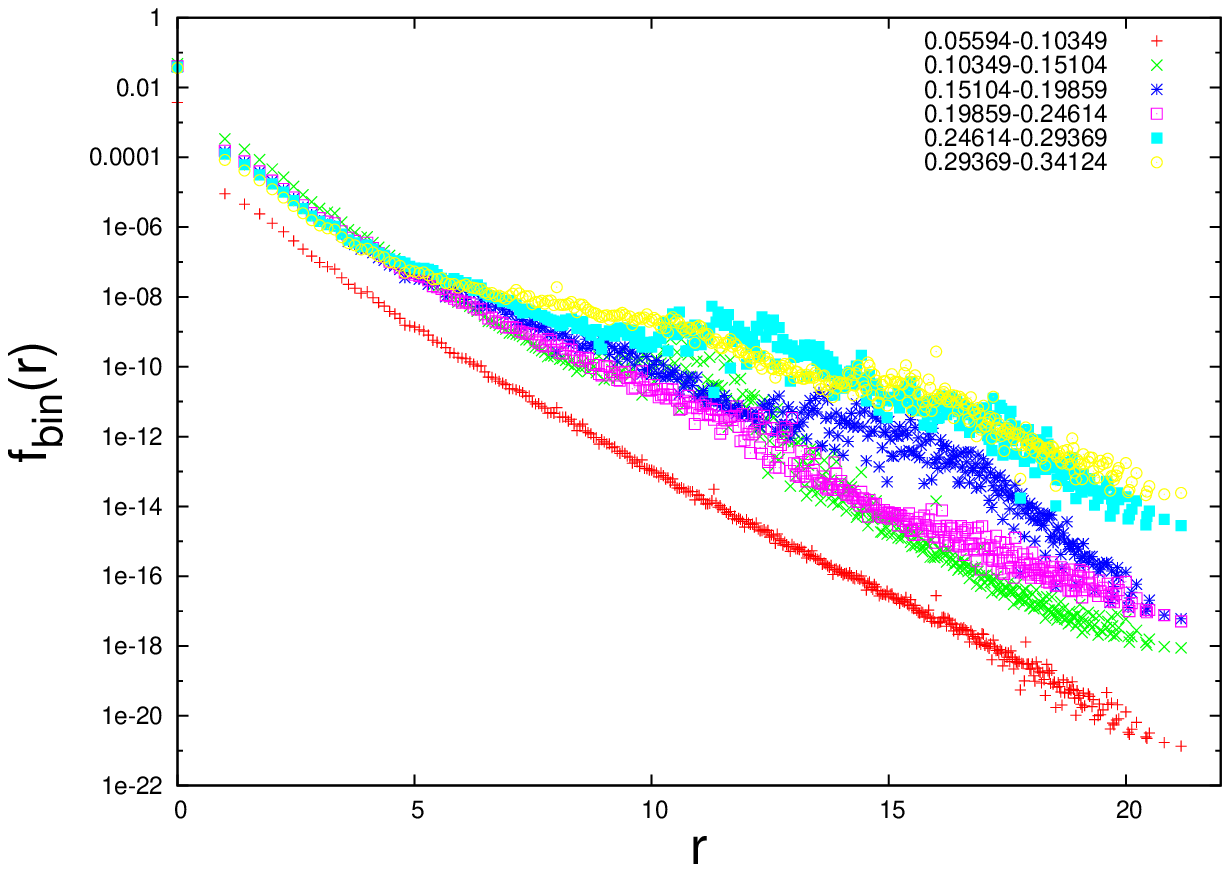}\\
    \includegraphics[scale=0.45]{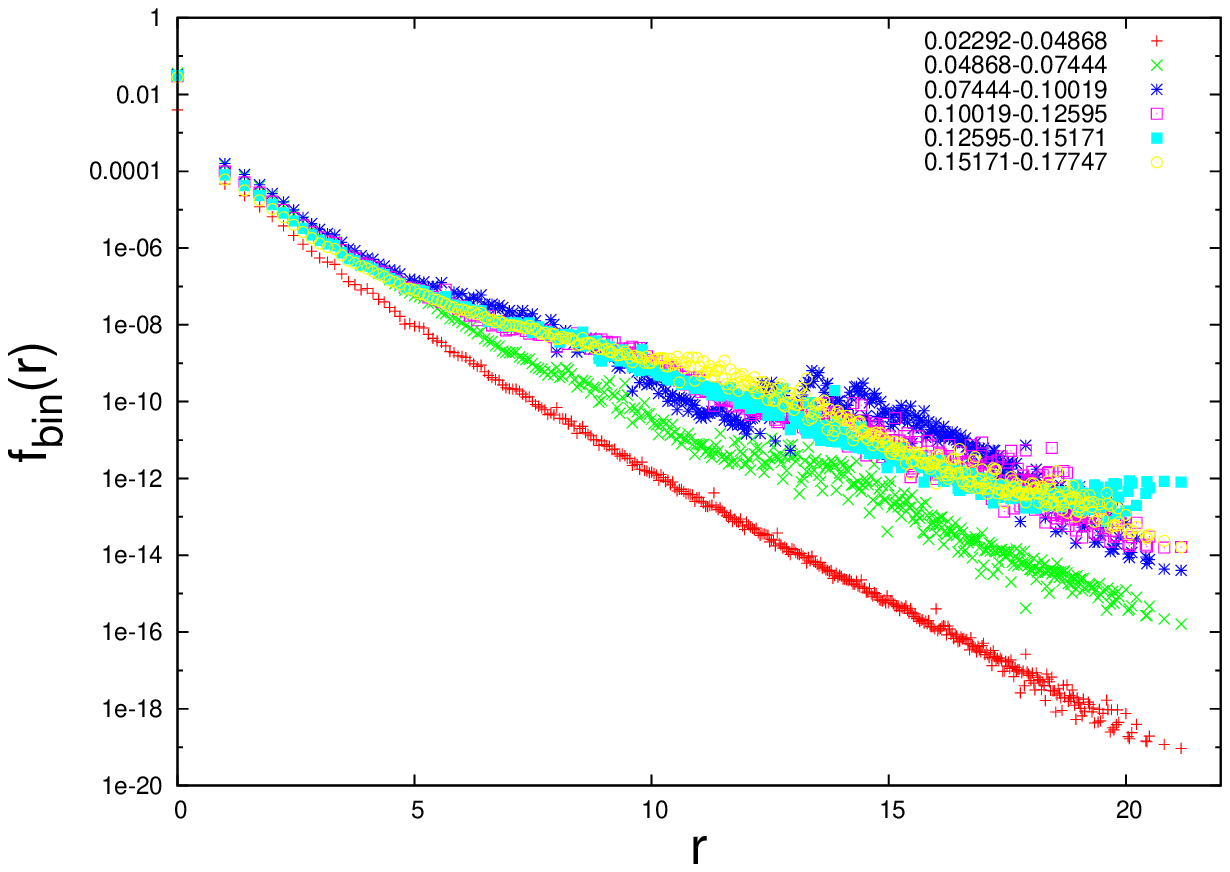}
    \includegraphics[scale=0.45]{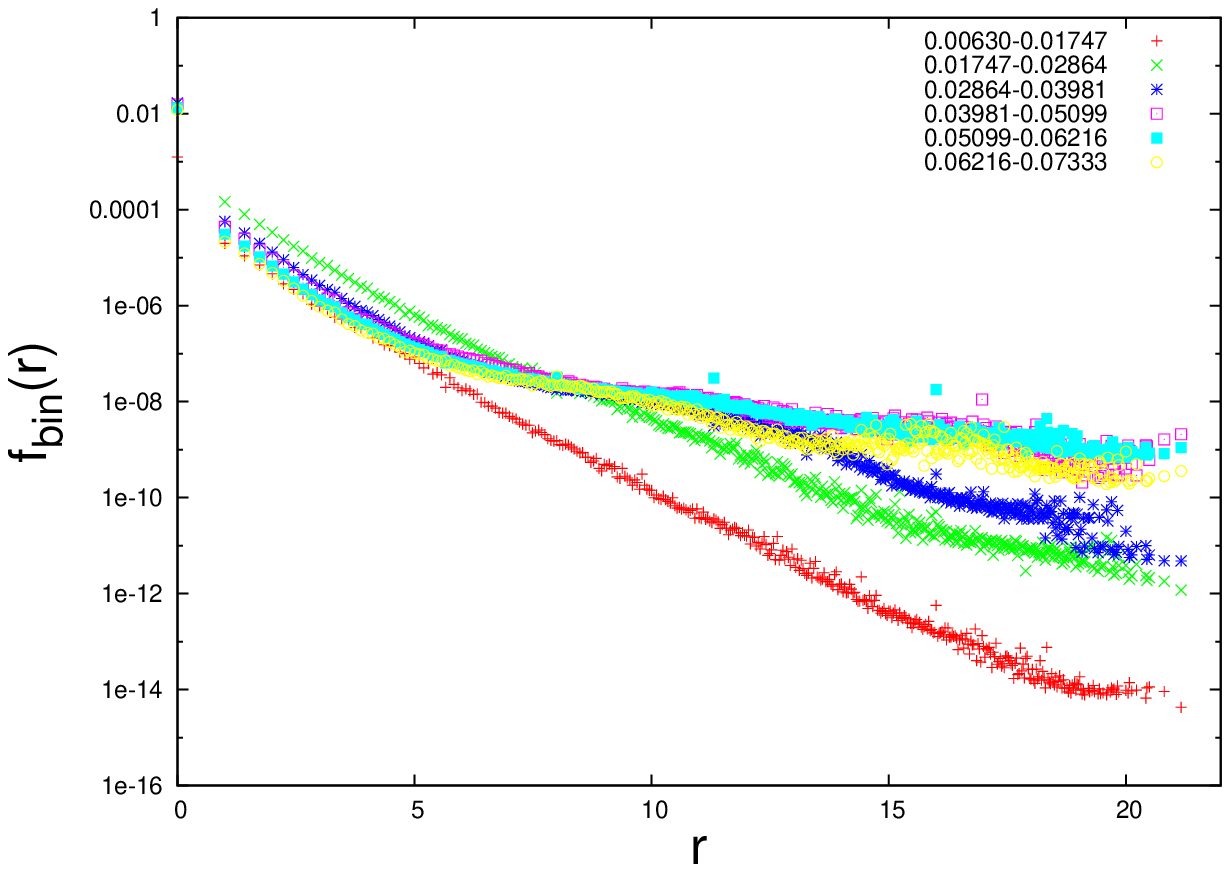}
  \end{center}
  \caption{
    Same as Figure~\protect\ref{fig2} but with the low-mode suppression term 
    ($\mu=0.2$).
    The data are shown for
    $\beta=5.78$ (top left), $\beta=5.68$ (top right),  
    $\beta=5.48$ (bottom left), and $\beta=5.28$ (bottom right).
  }
  \label{fig3}
\end{figure}

\begin{figure}[tb]
\begin{center}
	 \includegraphics[scale=0.45]{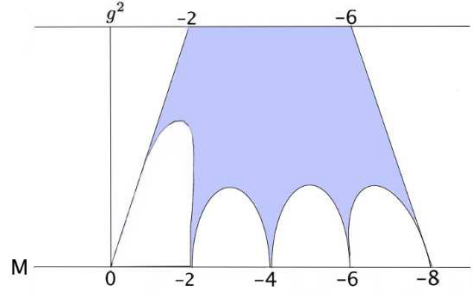}
\end{center}
\caption{
  Expected phase structure of the Wilson fermion after the low-mode
  suppression term is included.
}
\label{fig4}
\end{figure}

Once we introduce the low-mode suppression term, 
not only the eigenvalue spectrum but the profile of the eigenmode
changes as shown in Figure~\ref{fig3}.
Even at the $\beta$ value as low as 5.28, where the lattice spacing is
roughly 0.27~fm, the low-lying modes are still localized.
This implies that these parameter regions are outside of the Aoki phase. 
In other words, the Aoki phase structure is drastically changed by the
effect of the low-mode suppressing term.
The phase structure of the Wilson fermion would thus be changed as
illustrated in Figure~\ref{fig4}.

\section{Locality of overlap-Dirac operator }
The localization length of the overlap-Dirac operator $D_{ov}$ can
also be extracted directly from the behavior of $D_{ov}$.
We first set a source field as 
\begin{equation}
\eta_{\alpha}\left(x\right)=\begin{cases}
1 & x=(0,0,0,0)\\
0 & otherwise
\end{cases},
\end{equation}
where $\alpha$ is an index of internal degree of freedom for fermions.
Then we calculate a norm of a vector 
$\psi\left(x\right)=sgn\left(H_{W}\right)\eta\left(x\right)$ 
\begin{equation}
f\left(r\right)=\left\Vert \psi\left(x\right)\right\Vert ,\; r=\left\Vert x\right\Vert .
\end{equation}

\begin{figure}[H]
  \begin{center}
  	\includegraphics[scale=0.5]{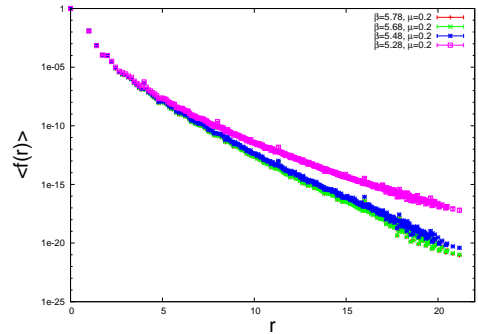}
  \end{center}
  \caption{$\langle f(r)\rangle$ as a function of $r$.
    Data for the low-mode suppressed gauge configurations: $\mu=0.2$.
  }
  \label{fig5}
\end{figure}

In Figure~\ref{fig5}, we plot
$\langle f(r)\rangle $ as a function of $r$
on the gauge configurations generated with the low-mode suppressing
term. 
For all four $\beta$ values calculated,
$\langle f(r)\rangle $ is rapidly decaying
as distance $r$ increases.

To determine the localization length, 
we fit $\left\langle f\left(r\right)\right\rangle$ to an exponential function
\begin{equation}
  \left\langle f\left(r\right)\right\rangle =
  c \exp\left(-r/l\right),
\end{equation}
at large distances $r$.
The results for the localization length $l$ is listed in
Table~\ref{table2}.
It shows that the overlap-Dirac operator can be properly defined at
strong couplings as far as the low-mode suppressing term is introduced.

\begin{table}[H]
\begin{center}
\begin{tabular}{|c|c|}
\hline 
$\beta$ & $l$ (localization length)\tabularnewline
\hline 
\hline 
5.78  & 0.58\tabularnewline
\hline 
5.68 & 0.57\tabularnewline
\hline 
5.48 & 0.60\tabularnewline
\hline 
5.28 & 0.76\tabularnewline
\hline 
\end{tabular}
\end{center}
\caption{Localizationion length calculated on the gauge configurations
  with the low-mode suppressing term $\mu=0.2$.}
\label{table2}
\end{table}

\section{Conclusion}
We studied the effect of the low-mode suppressing term on the locality
of the overlap-Dirac operator.
By inspecting the low-lying eigenmodes of $H_{W}$, we find the the
mobility edge is finite even at $\beta=5.28$, which corresponds to
$a\simeq 0.27$~fm.

\vspace*{1cm}
Numerical calculations are performed on Hitachi SR16000 at High Energy
Accelerator Research Organization (KEK) under a support of its 
Large Scale Simulation Program (No. 11-05).
SH is supported in part by the Grant-in-Aid of the Japanese Ministry
of Education (No. 21674002).

\bibliographystyle{unsrt}

\bibliography{pos}

\end{document}